\documentclass{aa_new}
\usepackage{psfig}
\usepackage{graphicx}

\begin{document}

\title{The first broad--band X--ray images and spectra of the 30~Doradus
       region in the LMC\,%
\thanks{Based on observations with XMM--Newton, an ESA Science Mission
        with instruments and contributions directly funded by ESA Member
        States and the USA (NASA)}
         }

\author{K. Dennerl\inst{1} \and
        F. Haberl\inst{1} \and
        B. Aschenbach\inst{1} \and
        U.~G.~Briel\inst{1} \and
        M. Balasini\inst{4} \and
        H. Br\"auninger\inst{1} \and
        W. Burkert\inst{1} \and
        R. Hartmann\inst{1} \and
        G. Hartner\inst{1} \and
        G. Hasinger\inst{3} \and
        J. Kemmer\inst{6} \and
        E. Kendziorra\inst{2} \and
        M. Kirsch\inst{2} \and
        N. Krause\inst{1} \and
        M. Kuster\inst{2} \and
        D. Lumb\inst{5} \and
        P. Massa\inst{4} \and
        N. Meidinger\inst{1} \and
        E. Pfeffermann\inst{1} \and
        W. Pietsch\inst{1} \and
        C. Reppin\inst{1} \and
        H. Soltau\inst{6} \and
        R. Staubert\inst{2} \and
        L. Str\"uder\inst{1} \and
        J. Tr\"umper\inst{1} \and
        M. Turner\inst{7} \and
        G. Villa\inst{4} \and
        V.~E. Zavlin\inst{1}
        }

\offprints{K. Dennerl, \email{kod@mpe.mpg.de}}

\institute{$^1$~Max-Planck-Institut f\"ur extraterrestrische Physik,
           85748 Garching, Germany \\
           $^2$~Institut f\"ur Astronomie und Astrophysik der Univ.\
           T\"ubingen, Waldh\"auserstr.~64, 
           72076 T\"ubingen, Germany \\
           $^3$~Astrophysikalisches Institut Potsdam, An der Sternwarte 16, 
           14482 Potsdam, Germany \\
           $^4$~Istituto di Fisica Cosmica ``G. Occhialini", Via Bassini 15, 
           20133 Milano, Italy \\
           $^5$~XMM Science Operations Centre, 
           Space Science Division ESTEC, Postbus 299, 2200\,AG
           Noordwijk, Netherland \\
           $^6$~KETEK GmbH, Am Isarbach 30, 85764 Oberschlei{\ss}heim, 
           Germany \\
           $^7$~Physics and Astronomy Department, University of Leicester, 
           Leicester LE1 7RH, United Kingdom \\
           }

\date{Received $<$date$>$ / Accepted $<$date$>$}

\abstract{
We present the XMM--Newton first light image, taken in January 2000 with the
EPIC pn camera during the instrument's commissioning phase, when XMM--Newton
was pointing towards the Large Magellanic Cloud (\object{LMC}). The field is
rich in different kinds of X--ray sources: point sources, supernova remnants
(SNRs) and diffuse X--ray emission from \object{LMC} interstellar gas. The
observations are of unprecedented sensitivity, reaching a few 10$\sp{32}$
erg/s for point sources in the \object{LMC}. We describe how these data sets
were analysed and discuss some of the spectroscopic results. For the SNR
\object{N157B} the power law spectrum is clearly steeper than previously
determined from ROSAT and ASCA data. The existence of a significant thermal
component is evident and suggests that \object{N157B} is not a Crab--like but a
composite SNR. Most puzzling is the spectrum of the \object{LMC} hot
interstellar medium, which indicates a significant overabundance of Ne and Mg
of a few times solar.
\keywords{Methods: data analysis --
          Techniques: image processing --
          ISM: abundances --
          ISM: supernova remnants --
          Galaxies: LMC --
          X--rays: ISM}
}

\maketitle



\section{Introduction}

On December 10 1999 the XMM--Newton spacecraft was placed in a 48 hour Earth
orbit by the first commercial ARIANE V launcher. XMM--Newton, or X--ray
Multi--Mirror Mission (Jansen et al.\ \cite{J}), is the second cornerstone of
the Horizon 2000 science program of the European Space Agency ESA.

The EPIC pn camera (Str\"uder et al.\ \cite{St}) was successfully commissioned
during the period of mid January through mid March (Briel et al.\ \cite{Br}).
On January 19 2000, shortly after the switch--on of the EPIC pn camera,
XMM--Newton received first light on the EPIC pn CCD chip. We selected as first
light target a region in the \object{LMC} about 10 arcmin southwest of the
\object{30 Doradus~A} complex. The field has been known to be very rich in
X--ray sources from Einstein (Long et al.\ \cite{L}) and ROSAT observations
(Tr\"umper et al.\ \cite{Tr}). There are at least three SNRs (one of those
known to contain a pulsar), one OB association and two R associations as well
as the \object{SN 1987A}. This XMM--Newton first light has already stimulated
detailed studies of some of the objects. Here we report some results of the
observation with four observation sequences of the same field added totaling
an exposure time of 106~ks. The observations are of unprecedented sensitivity,
reaching a few 10$\sp{32}$ erg/s for point sources. We present and discuss the
spectroscopic results over the broadest X--ray bandpass of 0.3\,--\,12~keV
ever covered by a single imaging X-ray instrument for the entire field of
27$\times$27 arcmin$^2$. In section 2 we describe details of the data analysis
techniques which may be useful for other observations as well. In section 3 we
discuss some of the scientific results.

\section{Observations and Data Reduction}

The region in the \object{LMC} was observed with the pn CCD camera operating
in full frame mode (frame time: 73~ms), with the medium and thin filter
inserted during the first three and last two observations, respectively
(Str\"uder et al.\ \cite{St} and Turner et al.\ \cite{Tu}). In Tab.\,\ref{obs}
we show the journal of the observations. For a fixed pointing direction some
photons fall onto the gaps between adjacent CCD chips. Fortunately, the
pointing direction and the roll angle were changed for the later three
observations with about the same exposure time as for the previous two
observations, such that the surface brightnesses are within a factor of 2,
with any noise contribution negligible. The data analysis was performed with
software which was developed for the analysis of ground calibration
measurements of the EPIC pn camera (Dennerl et al.\ \cite{De}). Most of this
software has been implemented into the XMM Standard Analysis Software (SAS,
Watson et al.\ \cite{Wat}).

\begin{table}[t]
\caption[]{Journal of Observations}
\begin{tabular}{lccc}
\hline\\[-3mm]
\multicolumn{1}{c}{Date} & Time & RA(2000) & DEC(2000) \\
\hline
\rule{0mm}{4mm}%
2000 Jan 19      & 16:19 -- 17:23  & 05 36 57.0 & -69 13 47 \\
2000 Jan 19/20   & 17:30 -- 04:28  & 05 36 57.0 & -69 13 47 \\
2000 Jan 21      & 15:38 -- 19:38  & 05 37 04.0 & -69 13 00 \\
2000 Jan 21/22   & 20:32 -- 07:01  & 05 37 04.0 & -69 13 00 \\
2000 Jan 22      & 09:08 -- 12:02  & 05 36 04.0 & -69 13 00 \\
\hline
\end{tabular}
\label{obs}
\end{table}

During the observations the background induced by low energy protons changed
and reached sometimes count rates of several hundred counts per second in the
full energy band. To enhance the signal to noise ratio, we screened the
observations for times with high background and excluded those times from
further processing. The net observing time after screening is 54.8~ks. In
addition the trails of minimum ionizing particles and reemission trails were
removed frame by frame. The remaining events were corrected for charge
transfer loss and gain variations, and events created by a single photon, but
split across several pixels, were recombined, to yield a photon event file in
detector coordinates with corrected energies (see Dennerl et al.\ \cite{De}
for a detailed description).

Test images produced from this data set in various energy bands indicated that
the energy range $E=0.3-5.0\mbox{ keV}$ was optimal to provide clean images
with high signal to noise ratio. The energy range was then split into three
narrower bands of similar signal to noise ratio, to create red, green and blue
components of the final rgb image. The variation in spectral hardness of the
LMC sources and the spectral resolution of the EPIC pn camera resulted in
colourful images which, however, do not represent the dynamic range in
intensity very well. To preserve the flux information as well, the colour
saturation was reduced by choosing overlapping energy ranges (0.3\,--\,1.0~keV
for red, 0.8\,--\,2.0~keV for green, and 1.5\,--\,5.0~keV for blue), and by
further distributing some of the flux from these bands into other bands.

\begin{figure}[ht]
\resizebox{\hsize}{!}{\psfig{file=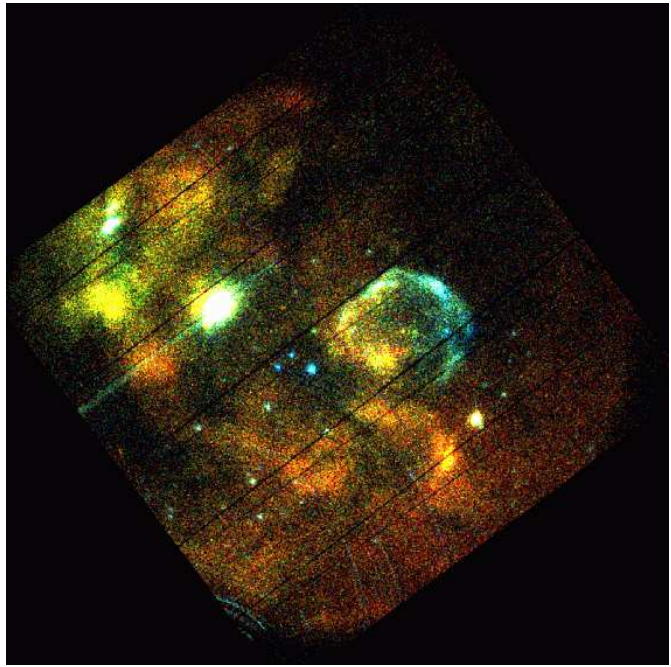,clip=}}
\caption{Superposition of the photons detected with the EPIC~pn camera in all
five \object{LMC} observations, displayed in equatorial coordinates. The
photon energy is coded in colour, ranging from 0.3~keV (red) to 5.0~keV
(blue). As this image has not been corrected for exposure variations, the CCD
boundaries are visible. The SNR \object{N157B}, the brightest source in the
image, is saturated; an out--of--time event trail extends along the CCD column
where the central neutron star is located. In the lower part of the image,
sharp blue arcs appear. They are caused by single mirror reflections of
photons from \object{LMC~X--1}, which is outside the FOV.}
\label{xmmlmcu}
\end{figure}

\begin{figure}
\resizebox{\hsize}{!}{\psfig{file=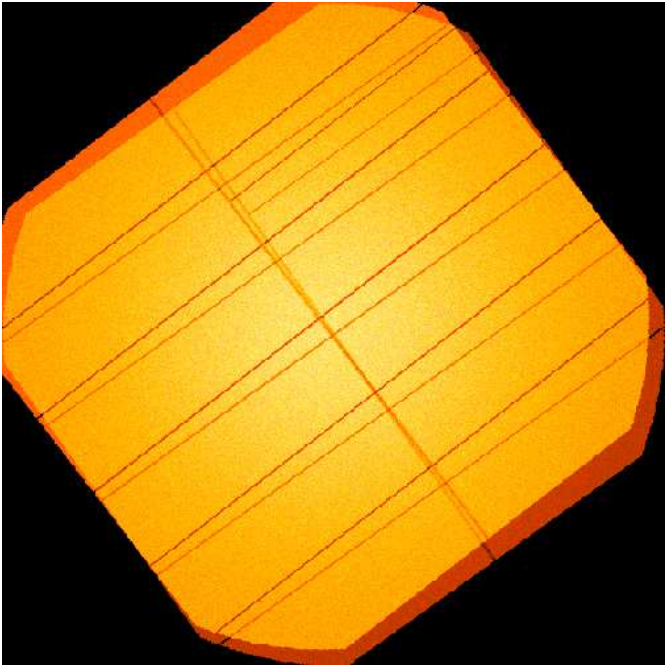,clip=}}
\caption{Combined exposure map of all five observations. For the final
image, only the inner part of the FOV was used, where the exposures overlap
completely. Since the pointing position and roll angle of XMM--Newton were
changed between the two sets of observations, almost all insensitive areas
of the detector were exposed in the complementary observation.}
\label{xmmlmce}
\end{figure}

True images of the sky were obtained by projection of each photon onto the
celestial plane, taking into account the different pointing directions and
roll angles. The projection takes care of the different spacing between the
individual CCDs as well as for the larger size of the pixels which are
furthest away from the readout nodes (Str\"uder et al.\ \cite{St}). The result
of the projection, after removing events from bright pixels and a bright
column, is shown in Fig.\,\ref{xmmlmcu}. The gaps between the CCDs show up in
the image, as no exposure correction was applied. The exposure map
(Fig.\,\ref{xmmlmce}) shows that these areas also received some light. The
exposure map was created in a similar way as the photon image, using the
vignetting curve for 1.5~keV from the XMM Users' Manual (Dahlem \cite{Da}).

\medskip
Apparent in Fig.\,\ref{xmmlmcu} is a streak--like feature from the SNR
\object{N157B}. This is an artefact of the operation of the CCD, which is
sensitive to X--rays also during readout. In the full frame mode, 7\,\% of all
events are recorded during this period and get a displaced position along
readout direction. This effect occurs every\-where on the detector, but
out--of--time event trails are evident only for bright sources. Suppressing
the trails does not only improve the image cosmetically, but also reduces
local variations of the background needed to apply the source detection
algorithms. In the following we describe a straightforward method to suppress
the contamination by out--of--time events.

\begin{figure}[ht]
\resizebox{\hsize}{!}{\psfig{file=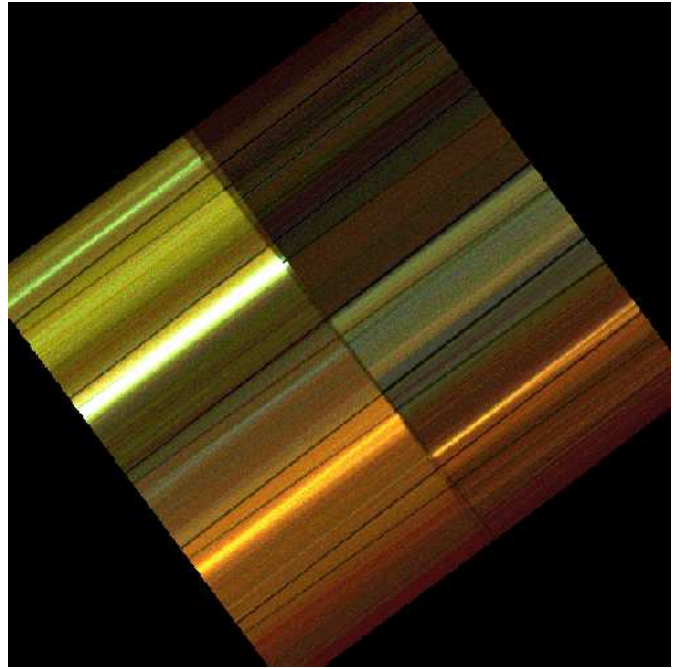,clip=}}
\caption{Out--of--time event image of Fig.\,\ref{xmmlmcu}, constructed by
distributing the events recorded in each readout column homogeneously across
the column. The intensity of this image was artificially increased with
respect to Fig.\,\ref{xmmlmcu} by a factor of 20 for better visibility of
these events.}
\label{xmmlmct}
\end{figure}

As the generation of out--of--time events is a random process, these events
become uniformly distributed along each readout channel. The fraction of
out--of--time events is given by the ratio of readout vs.\ frame time.
Thus, if $n$ is the total number of photons recorded in a particular column,
the contamination $n_{\rm cont}$ of any pixel in this column by out--of--time
events is
\[
n_{\rm cont} =  {t_{\rm read} \over t_{\rm frame} \cdot n_{\rm pix}} \cdot n
\]
where $t_{\rm read}$, $t_{\rm frame}$ and $n_{\rm pix}=200$ are the readout
time, the frame time and the number of pixels per column. For the full frame
mode, the values are $t_{\rm read}=5\mbox{ ms}$ and $t_{\rm frame}=73\mbox{
ms}$. Fig.\,\ref{xmmlmct} shows the contamination of Fig.\,\ref{xmmlmcu} by
out--of--time events, calculated separately for the three colour bands in
detector coordinates and then projected onto the sky, so that it can be
subtracted from Fig.\,\ref{xmmlmcu}.

\begin{figure}[ht]
\resizebox{\hsize}{!}{\psfig{file=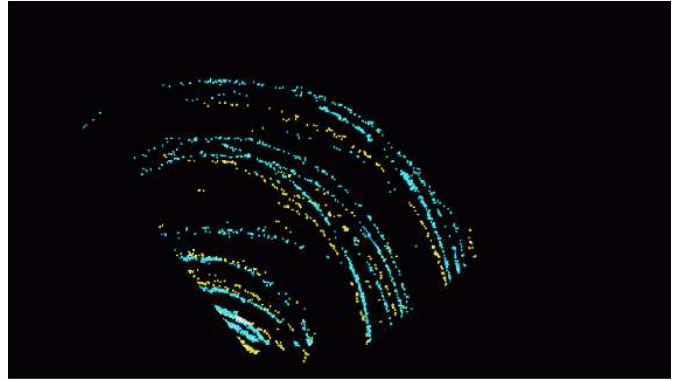,clip=}}
\caption{Single mirror reflections of \object{LMC~X--1} in the lower corner of
Fig.\,\ref{xmmlmcu}. Pixels affected during the first two observations are
marked in yellow, while pixels affected during the last three observations
are coded in blue. The correction was done individually for the two sets of
observations.}
\label{xmmlmcr}
\end{figure}

\begin{figure*}
\resizebox{\hsize}{!}{\psfig{file=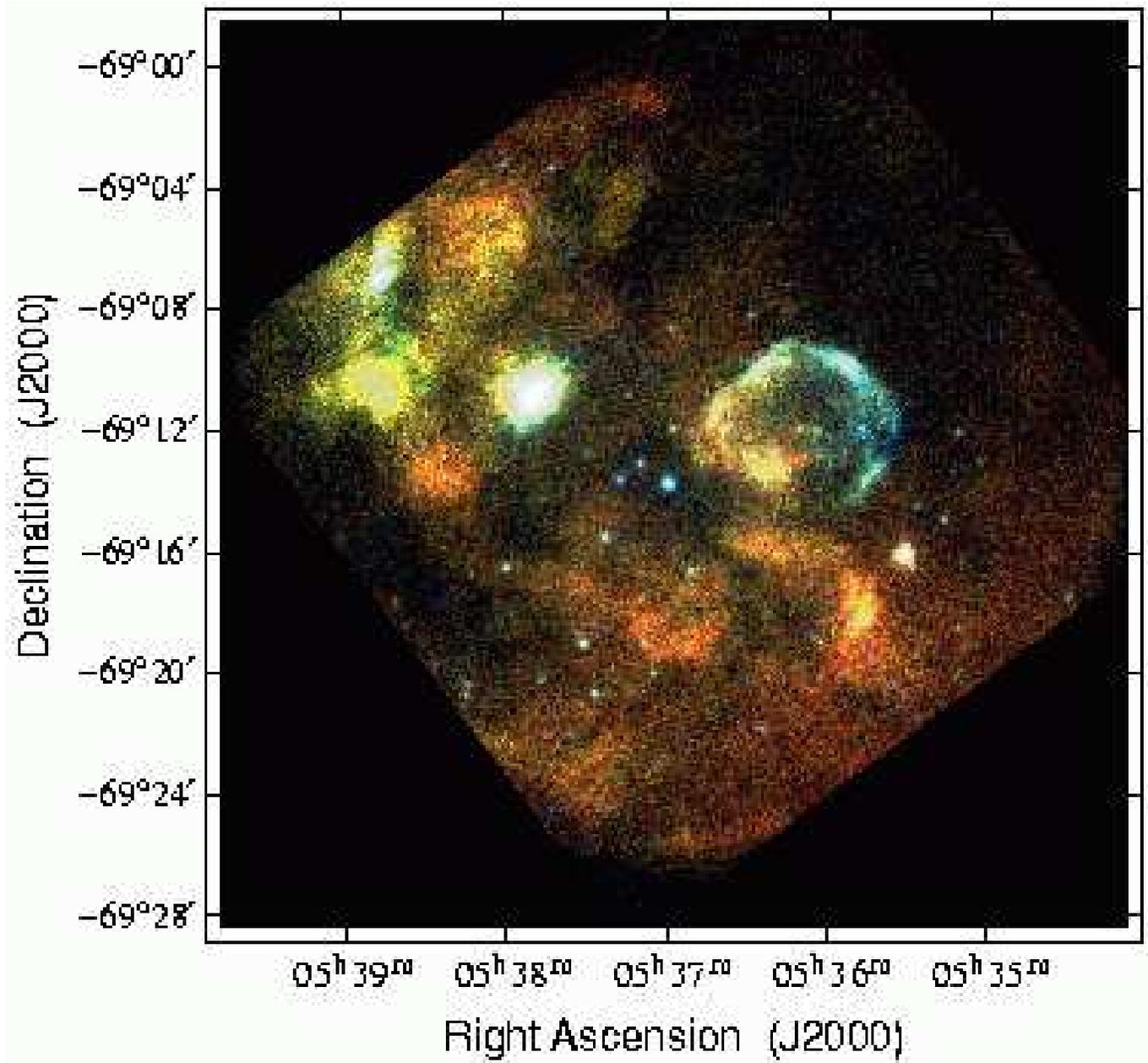,clip=}}
\caption{Final version of the XMM--Newton EPIC pn first light X--ray colour
image of the \object{LMC}, after applying all the data reductions described in
the text.}
\label{xmmlmc}
\end{figure*}

An additional source of undesired `background' in Fig.\,\ref{xmmlmcu} are
sharp, blue arcs in the lower corner, which would systematically contaminate
the spectra of the diffuse soft X--ray emission in this region. These are
caused by single mirror reflections of photons from the bright X--ray binary
\object{LMC~X--1}, which is outside the FOV. In order to suppress these arcs,
we identified the contaminated pixels from the hard images, where they are most
distinct, and flagged them (Fig.\,\ref{xmmlmcr}). For the first light image,
we replaced the flux of the affected pixels by interpolated values from the
adjacent pixels. This approach seems to be justified, since the extent of one
pixel ($4\farcs13$) is considerably smaller than the half energy width of the
telescope ($15\farcs1$), which implies that the fluxes of adjacent pixels
cannot be independent of each other due to the point spread function of the
telescope.

\medskip
A final challenge for producing the first light image was the high dynamic
range in intensity, which exceeds three orders of magnitude between the
diffuse extended regions of low surface brightness and the point source in
\object{N157B}. Since standard techniques like logarithmic intensity coding
did not yield satisfying results, we designed a special code which preserves
different shades of gray for the brighest sources. The final image, boxed with
an equatorial coordinate grid, is shown in Fig.\,\ref{xmmlmc}.

\section{Results and Discussion}

As Fig.\,\ref{xmmlmc} shows, there is a wealth of different objects,
point--like and extended up to 130~pc in size. The X--ray colours show large
spectral differences from object to object. Based on this image, we
accumulated photons from selected regions and performed spectral
investigations of various sources, to utilize the full resolving power of the
EPIC pn camera. We did this separately for the three data sets which were
taken with the same observational setup (filter and pointing direction),
because they require different detector response matrices. The best fit was
obtained by modeling each of the three count rate spectra with the same
candidate source spectrum but with the individual normalization factor left as
a free parameter (for details see Haberl et al.\ \cite{Hab}). Errors were
determined at $\Delta\chi^2=+2.3$ with all parameters free.

\subsection{\object{30 Dor C}}

The largest object in the field is \object{30 Dor C}, already known from radio
observations and suggested by Mills et al.\ (\cite{Mi}) to be an SNR. In
X--rays it appears as a complete ring of $\sim6$~arcmin diameter. The size of
the shell and its colour make this remnant the first of its kind outside our
Galaxy. In a separate paper (Aschenbach et al.\ \cite{A1}) this unique object
is investigated in detail. Close to the geometric center we detect a
previously unknown point source, displayed in yellow, which emits mainly in
the 0.3\,--\,1.0~keV band. If this source is in the \object{LMC}, then its
luminosity is of the order of $10^{33}\mbox{ erg s}^{-1}$, and it might be the
compact remnant of \object{30 Dor C}, either a neutron star or a black hole.

\subsection{\object{SN 1987A}}

Approximately 2~arcmin to the south-west of the shell border is the well--known
\object{SN 1987A}. The XMM--Newton observations are very important with
respect to both, the light curve and the X--ray spectrum, refining current
models of \object{SN 1987A}. Both aspects are presented and discussed in an
accompanying paper by Aschenbach et al.\ (\cite{A2}).

\subsection{Active Galactic Nuclei behind the \object{LMC}}

A number of point sources have been discovered, displayed as blue dots in
Fig.\,\ref{xmmlmc}, the brightest of which is seen close to the image center.
These sources have power law spectra and show strong low energy absorption.
Both, the spectra and the fact that the central source is associated with a
radio point source, indicate that we see background sources, probably AGN,
Nuclei, through and far beyond the \object{LMC}. A detailed analysis of these
new X--ray sources is presented in an accompanying paper by Haberl et al.\
(\cite{Hab}).

\subsection{\object{N157B}}

\begin{figure}[tbp]
\resizebox{\hsize}{!}{\psfig{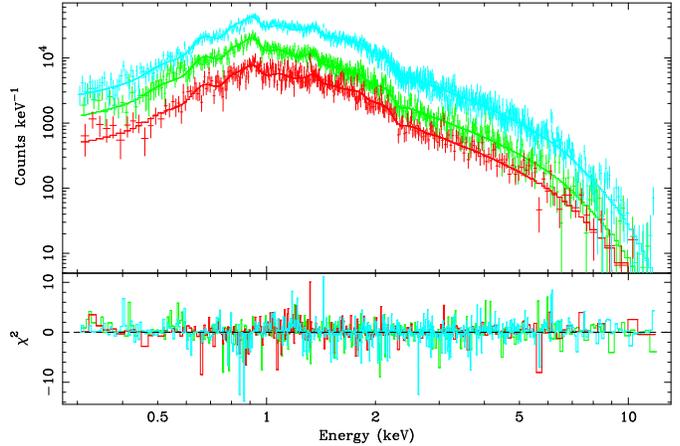}}
\caption[ ]{EPIC pn X--ray spectra of N157B for the three data sets with
the same observational setup. The histograms represent the best fit model
simultaneously applied to the spectra.}
\label{n157b}
\end{figure}

By far the brightest object in the field is the SNR \object{N157B}, located
halfway from the image center to the edge in north--easterly direction. After
the discovery of the 16~ms pulsar (Marshall et al. \cite{Ma}, Wang \& Gotthelf
\cite{Wa2}) \object{N157B} has been classified as a Crab--like remnant with a
central pulsar and a surrounding synchrotron nebula. In a preceeding paper
Wang \& Gotthelf (\cite{Wa1}) showed that the spectrum is dominated by a power
law with an indication of some thermal emission. The XMM--Newton spectrum of
\object{N157B} (Fig.\,\ref{n157b}) demonstrates convincingly for the first
time the presence of emission lines at low energies, proving the presence of a
thermal component. A model comprising two thermal components and a power law
gives an acceptable fit to the data. Formally, ionization equilibrium models
like Raymond--Smith (RS; Raymond \& Smith \cite{R}) or non equilibrium
ionization models (NEI; Hamilton et al.\ \cite{Ham}) of XSPEC used for the
thermal components at low energies cannot be discriminated.

Fig.\,\ref{n157b} shows the best fit two temperature RS models with a power
law model added. The temperatures are 0.37$^{+0.11}_{-0.05}$ keV and
0.12$\pm0.01$ keV, and the photon index of the power law is 2.83$\pm$0.03.
With an assumed hydrogen equivalent photoelectric absorption of
$6\cdot10^{20}$ H--atoms cm$^{-2}$ for the galactic foreground, we derive an
additional H--equivalent column density of (1.94$\pm$0.04)$\cdot10^{22}$
H--atoms cm$^{-2}$ for the absorption in the \object{LMC}, taking elemental
abundances of 0.5 solar into account. To fit the Mg emission line at 1.3~keV
clearly seen in the spectra, an overabundance of 1.3 solar is required (for
the cooler component). The NEI model yields similar values for the absorption,
the Mg abundance and the power law slope, but different temperatures (0.24 and
0.10 keV).

The existence of the additional two--temperature thermal component suggests
that \object{N157B} is not a Crab--like remnant but a composite remnant with an
X-ray bright pulsar and synchrotron nebula but relatively faint thermal shell.
The analysis of the thermal component, to be done in a separate paper, will be
used to check for instance the consistency of the age derived from a
Sedov--type approach with the pulsar characteristic spin-down age of
$P/(2\,\dot P)\sim5000\mbox{ ys}$ (Wang \& Gotthelf \cite{Wa2}).

The power law component in the spectrum of \object{N157B} which has been
discovered by Wang \& Gotthelf (\cite{Wa1}) using ROSAT PSPC and ASCA SIS data
gives a photon index of 2.3$\pm$0.1. The EPIC pn best fit value for the index
is 2.83$\pm$0.03. \object{N157B} has been compared with the \object{Crab}.
Both the \object{Crab pulsar} and the \object{N157B 16~ms pulsar} (Marshall et
al. \cite{Ma}) show a flat power law spectrum with a photon index of 1.5 and
1.6$\pm$0.3, respectively. The spectrum of the \object{Crab nebula} is steeper
by 0.5, generally believed to be due to synchrotron losses. If the spectrum of
\object{N157B} would have a similar difference of $\sim0.5$ between the pulsar
and the nebula spectrum as suggested by the analysis of Wang \& Gotthelf, the
analogy with the \object{Crab} would be convincing. If on the other hand the
EPIC pn power law slope is a better representation of the spectrum, the
analogy with the \object{Crab} is cleary violated and other acceleration and
loss processes have to be considered. We note that the analysis of power law
dominated spectra like in AGN show just minor discrepancies in the spectral
index between ASCA and EPIC pn. Because of the steep spectrum of the nebula
the physical conditions, either the electron injection spectrum, the electron
transport speed or the magnetic field distribution, are likely to be different
from those prevailing in the \object{Crab}.

\subsection{The \object{Honeycomb nebula}}
\begin{figure}[ht]
\resizebox{\hsize}{!}{\psfig{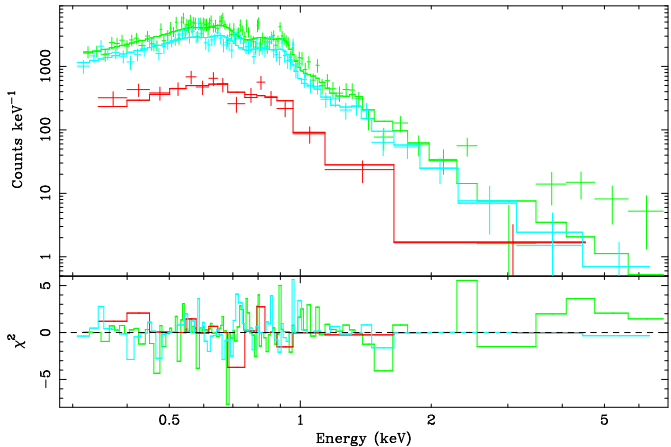}}
\caption[]{Same as Fig.\,\ref{n157b} for the \object{Honeycomb nebula}.}
\label{honey}
\end{figure}

In Fig.\,\ref{xmmlmc}, X--ray emission from the \object{Honeycomb nebula} is
seen 1~arcmin south--east of \object{SN 1987A}. The emission of the bright
orange patch can be modeled by a thermal plasma with a temperature of
0.19$\pm$0.02~keV, superimposed on a power law of index 3.6$\pm$0.6. Increased
Ne and Mg abundances ($\sim$2 times solar) are indicated. The hydrogen
equivalent LMC absorption to this SNR is (2.2$\pm$0.9)$\cdot10^{21}$ H--atoms
cm$^{-2}$. Fig.\,\ref{honey} shows the three EPIC pn spectra together with the
best fit model.

\begin{figure}[]
\resizebox{\hsize}{!}{\psfig{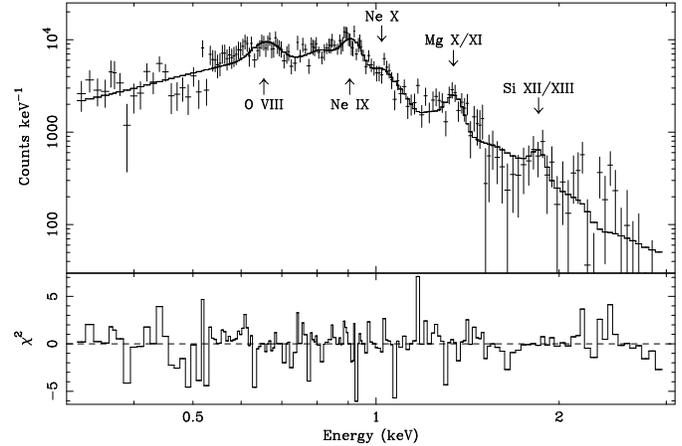}}
\caption[ ]{EPIC pn X--ray spectrum of the faint, diffuse emission in the
north of Fig.\,\ref{xmmlmc}, accumulated from the last two observations
with thin filter.}
\label{s00}
\end{figure}

\subsection{Interstellar medium in the \object{LMC}}

The north--western part of Fig.\,\ref{xmmlmc} appears remarkably dark and
empty. But when we extract a spectrum from a 44~arcmin$^2$ region north of
\object{30 Dor C} where no point sources and no out--of--time events from
\object{30 Dor C} affect the data, we clearly see thermal emission. The
temperature appears to differ from place to place over a range from $\sim$0.1
keV and $\sim$0.3 keV, as was already concluded from ROSAT PSPC spectra
(Sasaki et al.\ \cite{Sa}). But for the first time we detect emission lines of
oxygen, neon, magnesium and silicon (Fig.\,\ref{s00}). To fit the spectrum, we
subtracted the background in a preliminary way, using source free areas from
other EPIC pn observations. The resulting spectrum is well represented by a
0.3$^{+0.05}_{-0.2}$ keV RS type model with overabundances in O, Ne, Mg and Si
of 1.6$\pm$0.7, 5.4$\pm$1.5, 6.6$\pm$2.5 and 8.9$\pm$7.4 solar, respectively,
which is surprising in view of the generally low metallicity of the LMC.

\begin{acknowledgements}
The XMM--Newton project is supported by the Bundesministerium f\"ur Bildung
und Forschung\,/\,Deutsches Zentrum f\"ur Luft- und Raumfahrt (BMFT/DLR),
the Max--Planck Society and the Heidenhain--Stiftung.
\end{acknowledgements}


\end{document}